\documentclass[twocolumn,showpacs,amsmath,amssymb,superscriptaddress,aps,prc]{revtex4-1}

\usepackage[dvips]{graphicx}
\usepackage{epsfig}
\usepackage{amsmath}
\usepackage[dvips]{color}
\newcommand{\beq}{\begin{eqnarray}}
\newcommand{\eeq}{\end{eqnarray}}
\newcommand{\be}{\mbox{$^{9}$Be}}

\newcommand{\bepb}{\mbox{$^{9}$Be+$^{208}$Pb}}
\newcommand{\beal}{\mbox{$^{9}$Be+$^{27}$Al}}

\newcommand{\emax}{\mbox{$E_{\rm max}$}}

\newcommand{\jmax}{\mbox{$j_{\rm max}$}}

\begin{document}
	\title{$\be$ scattering with microscopic wave functions and the CDCC method}
	\author{P. Descouvemont}
	\email{pdesc@ulb.ac.be}
\affiliation{Physique Nucl\'eaire Th\'eorique et Physique Math\'ematique, C.P. 229,
	Universit\'e Libre de Bruxelles (ULB), B 1050 Brussels, Belgium}
\author{N. Itagaki}
\email{itagaki@yukawa.kyoto-u.ac.jp}
\affiliation{Yukawa Institute for Theoretical Physics, Kyoto University, Kitashirakawa Oiwake-Cho, Kyoto 606-8502, Japan}
\date{\today}
\begin{abstract}
We use microscopic $\be$ wave functions defined in a $\alpha+\alpha+n$ multicluster model to compute
$\be$+target scattering cross sections. The parameter sets describing $\be$ are generated in the
spirit of the Stochastic Variational Method (SVM), and the optimal solution is obtained
by superposing Slater determinants and by diagonalizing the Hamiltonian. The $\be$ three-body continuum is approximated
by square-integral wave functions. The $\be$ microscopic wave functions are then used in a
Continuum Discretized Coupled Channel (CDCC) calculation of $\bepb$ and of $\beal$ elastic scattering. Without any
parameter fitting, we obtain a fair agreement with experiment. For a heavy target, the influence of $\be$
breakup is important, while it is weaker for light targets. This result confirms previous non-microscopic
CDCC calculations. One of the main advantages of the microscopic CDCC is that it is based on nucleon-target
interactions only; there is no adjustable parameter. The present work represents a
first step towards more ambitious calculations involving heavier Be isotopes.
\end{abstract}
\maketitle

\section{Introduction}
Exotic nuclei represent a major interest in current nuclear physics \cite{TSK13}.  These nuclei, close to the 
driplines, are characterized by a low binding energy of the last nucleon(s).  This property leads to a halo 
structure, well known since 30 years \cite{THH85}.  A halo nucleus is considered as a core surrounded 
by one or two nucleons.  Owing to the low binding energy, the spatial extension of the valence nucleons is large, 
and the associated radii are much larger than in stable nuclei.  Exotic nuclei present further 
interesting properties such as a change of the magic numbers \cite{OKS00}. 

Most exotic nuclei have a short lifetime, and can be investigated by reactions only.  The recent development 
of radioactive beam facilities over the world provided many new data, which require more and more 
sophisticated models.  The weak binding energy of the projectile, however, needs a special attention 
since it strongly affects the various cross sections (elastic scattering, breakup, fusion, etc.).  
To address this issue, a well known approach is the Continuum Discretized Coupled Channel (CDCC) method 
\cite{AIK87} which was originally developed to investigate deuteron scattering \cite{Ra74}.

The CDCC method is based on a structure model for the projectile.  The simplest approach is of course a 
two-body model, where the projectile consists of two structureless clusters.  Typical examples 
are $d=p+n$, $^{11}{\rm Be} = ^{10}$${\rm Be} +n$ or $^{8}{\rm B} = ^{7}$${\rm Be}+p$.  Many works have been 
performed within this approach.  For some nuclei, however, this two-body description is not adapted, and 
extensions have been recently developed.  The first extension is a three-body model of the projectile, 
which is necessary for nuclei such as $^{6}{\rm He}$ \cite{MHO04}, $^{9}{\rm Be}$  \cite{DDC15,CRA15} 
or $^{11}{\rm Li}$  \cite{CFR12}.  Another development of CDCC involves core excitations which 
are considered in $^{19}{\rm C}$  \cite{LDC16} and in $^{11}{\rm Be}$  \cite{DCM17}, for example.

The determination of the scattering matrices in the CDCC method makes use of fragment-target optical 
potentials.  This technique provides a set of coupled-channel equations which leads to the projectile-target wave 
functions and to the scattering matrices.  The projectile breakup, important for weakly bound nuclei, 
is simulated by approximate continuum states of the projectile.  These states, also referred to as 
pseudostates, do not have a physical meaning, but allow to take account of the projectile breakup.  
For weakly bound nuclei, the cross sections are in general different whether breakup is included 
or not.  As mentioned before, most CDCC calculations are performed within a two-body or a three-body model.  A drawback 
of this approach is that it requires the knowledge of fragment-target optical potentials, which 
are sometimes poorly known, or not known at all.  In these circumstances, simplifying assumptions are necessary.  

In a recent development, the projectile is described within a microscopic model, where all nucleons are 
involved \cite{DH13,DPH15}.  The main advantage of this approach is that only nucleon-target optical 
potentials are needed.  These potentials are known over a wide range of energies and masses, and 
accurate parametrizations are available \cite{VTM91,KD03}.  Several variants of microscopic models have
been developed.  In particular, cluster models \cite{HIK12} are well adapted to CDCC calculations.  
In cluster models, the $A$-nucleon structure is taken into account, but the nucleons are assumed 
to be grouped in clusters \cite{WT77}.  This approximation permits a simplification of the calculations, 
whilst it keeps the microscopic character of the model.  Besides, the cluster structure is a natural starting point 
for the treatment of breakup.  Recently, $^{7}{\rm Li}$ \cite{DH13} and $^{6}{\rm He}$ \cite{De16b} scatterings 
have been studied within this formalism.

A challenge for cluster models is the description of nuclei involving several clusters.  In that case, 
the number of degrees of freedom is large, and the choice of the basis functions requires a special 
attention.  This problem can be efficiently addressed by using the Stochastic Variational Method (SVM) \cite{KK77,VS95}. 
The SVM randomly generates parameter sets of the wave function, and it permits to achieve 
convergence by superposing several Slater determinants, even when many parameters are involved (see, for example, Ref.\ \cite{HS14}).  
Our goal for the future is to apply the SVM to nucleus-nucleus reactions, where one of the colliding 
particles is an exotic light nucleus.  In particular, $^{11}{\rm Be}$+target reactions provide a strong 
evidence for a halo structure in $^{11}{\rm Be}$ \cite{DSM12}.  The traditional description of $^{11}{\rm Be}$ 
is a two-body $^{10}{\rm Be} + n$ potential model.  However, our aim is to go beyond this simple approximation, 
and to use a microscopic description of $^{11}{\rm Be}$.  It has been shown that a multicluster 
microscopic model based on two $\alpha$ particles and on additional neutrons  provides a precise description of low-lying states of Be isotopes \cite{PhysRevC.61.044306,PhysRevC.62.034301,PhysRevC.65.044302,PhysRevC.66.057301,PhysRevC.68.054302}.  

In the present work, we want to explore multicluster wave functions for the projectile description.  Our first 
application deals with $\be$, considered as an $\alpha+\alpha+n$ three-cluster system.  
The microscopic structure calculation is performed using the idea of SVM, and obtained
transition densities are utilized in the CDCC calculation. It is known that breakup effects are likely weaker in
$^{9}{\rm Be}$ and $^{10}{\rm Be}$ than in $^{11}{\rm Be}$ \cite{DSM12}.
However, many data on 
elastic scattering are available, and $^{9}{\rm Be}$ + target scattering is an excellent test before 
considering more ambitious systems, involving $^{10}{\rm Be}$ or $^{11}{\rm Be}$. As the model is parameter free, validity tests on well known systems are necessary.

The paper is organized as follows.  In Sec.\ \ref{sec2}, we present the microscopic model of $\be$, 
and discuss the main properties (energy spectrum, r.m.s. radii, electric transition probabilities, etc.).  
Section \ref{sec3} is devoted to a brief outline of the microscopic CDCC method.  We apply the 
scattering model to $\beal$ and $\bepb$ systems in Sec.\ \ref{sec4}.  These reactions involve a light target, 
$^{27}{\rm Al}$, and an heavy target, $^{208}{\rm Pb} $.  Concluding remarks and outlook are 
presented in Sec.\ \ref{sec5}.

\section{Microscopic description of the $\be$ structure}
\label{sec2}

In this section, we explain the structure calculation,
which is the microscopic description of $^9$Be based on the $\alpha$+$\alpha$+$n$ model.
The idea of the SVM is used for the generation of the basis states.
Additional information can be found in Refs.~\cite{PhysRevC.66.057301,PhysRevC.90.039902}.
From the wave functions we determine the transition densities, which are then used in the CDCC calculations.

\subsection{$\be$ Hamiltonian}

In a microscopic approach, the $\be$ Hamiltonian $H_0$ depends on all nucleon coordinates, and is given by 
\begin{equation}
H_0=\sum_{i=1}^{9}t_i- T_{c.m.} + \sum_{i\leq j=1}^{9}v_{ij},
\label{eq_h0}
\end{equation}
where the center-of-mass kinetic energy $T_{c.m.}$ is subtracted to guarantee the
translation-invariance of the wave functions.
In this Equation, $t_i$ is the kinetic energy of nucleon $i$, and $v_{ij}$ a 
nucleon-nucleon interaction. 
The two-body nucleon-nucleon interaction $v_{ij}$ consists of
central ($v_{ij}^{\rm central}$), spin-orbit ($v_{ij}^{\rm spin-orbit}$), 
and Coulomb parts.

For the central interaction, we adopt the Minnesota potential~\cite{THOMPSON197753}
which involves the exchange parameter $u$. The standard value is $u=1$, but it can be slightly modified to reproduce important properties of the system. We have used different values for both parities, in order to reproduce the experimental binding energies of the $3/2^-$ ground
state and of the $1/2^+$ first excited state ($-1.57$ MeV and 0.11 MeV, respectively).  These constraints provide $u=0.993$ for positive parity, and $u=0.967$ for negative
parity. Throughout the text, energies are defined with respect to the $\alpha+\alpha+n$
threshold.

For the spin-orbit part, we adopt a one-Gaussian type 
interaction
\begin{equation}
v_{ij}^{\rm spin-orbit}= V_{ls} \pmb{L}\cdot \pmb{S}\, \exp\bigl[-(\pmb{r}_i - \pmb{r}_j)^2/r_{ls}^2\bigr]/r_{ls}^5,
\label{Vls}
\end{equation}
where the operator $\pmb{L}$ stands for the relative angular momentum,
and where $\pmb{S}$ is the total spin, ($\pmb{S} = \pmb{S}_{1}+\pmb{S}_{2}$).
The strength and range,
$V_{ls} = -20$ MeV.fm$^5$ and $r_{ls} = 0.1$ fm, 
have been tested in many previous cases, and we adopt these values. The Coulomb interaction
is treated exactly.

For given spin $j$ and parity $\pi$, Hamiltonian (\ref{eq_h0}) is then diagonalized as
\beq
H_0\, \Phi^{jm\pi}_{k}=E^{j\pi}_{0,k} \, \Phi^{jm\pi}_{k},
\label{eq2}
\eeq
where $k$ is the excitation level.  
In the CDCC framework, negative energies $E^{j\pi}_{0,k}$ correspond to physical states, and positive energies to pseudostates, 
which can be considered as discrete approximations of the continuum.
Different techniques are used to find approximate solutions of 
(\ref{eq2}): the Resonating Group Method \cite{Ho77}, the Antisymmetrized Molecular Dynamics \cite{KH01}, 
the Fermionic Molecular Dynamics \cite{FS00} or the Molecular Orbit Model \cite{IH02} are typical methods.

For the CDCC calculation, we need the transition densities.
They are calculated with the wave functions of the static calculation as
\begin{equation}
\rho_{kl}^{j_1m_1,j_2m_2} (\pmb r)  =   \langle \Phi^{j_1m_1}_{k} | \sum_i 
\bigl(\frac{1}{2}\pm t_{iz}\bigr)\delta (\pmb r - \pmb r_i) |  \Phi^{j_2m_2}_{l}  \rangle, 
\label{rho_1}
\end{equation}
where $i$ runs over protons or neutrons. In this definition, $\pmb{t}_i$ is the isposin of
nucleon $i$, and the signs "+" and "-" correspond to the neutron and proton densities,
respectively.
In the actual calculation, we write the density in a multipole expansion \cite{Ka81} as
\begin{eqnarray}
\rho_{kl}^{j_1m_1,j_2m_2} (\pmb r)   &=& \sum_{\lambda } \langle j_2m_2  \lambda m_1-m_2 | j_1m_1 \rangle \nonumber \\
&&\times \rho^{j_1,j_2}_{kl,\lambda}(r) 
Y_{\lambda m_1-m_2}^*(\Omega_r)
\label{rho_2}
\end{eqnarray}
and calculate the matrix elements of $\rho^{j_1,j_2}_{kl,\lambda}(r)  $.
For the purpose of applying to reaction calculations,
we have to carefully describe the tail regions of the densities.
The method to directly calculate the multipole densities can be found in Ref.~\cite{BDT94},
and we adopt the same formalism.

\subsection{Basis wave functions}
A $\be$ intrinsic wave function $\Psi$ is the antisymmetrized product of 
single-particle wave functions $\psi_n$ as
\begin{equation}
\Psi = {\cal A} \{ \psi_1 \psi_2 \psi_3 \cdot \cdot \cdot \cdot \psi_9 \},
\label{total-wf}
\end{equation}  
where $\psi_n$ has a Gaussian shape
\begin{equation}
        \psi_{n}(\pmb{r}_n) = \left( \frac{2\nu}{\pi} \right)^{\frac{3}{4}}
        \exp \left[- \nu \left(\pmb{r}_n - \pmb{R}_n \right)^{2} \right] \eta_{n}.
\label{spwf}
\end{equation}
In this definition, $\eta_{n}$ represents the spin-isospin component of the wave function,
and $\pmb{R}_n$ is a parameter representing the center of a Gaussian
function for the $n$-th nucleon.
The size parameter $\nu$ is equal to $\nu=1/2b^2$ and the oscillator parameter $b$ is chosen as
1.36 fm, a standard value for the $\alpha$ particle.

Based on the generator coordinate method (GCM), the superposition of
different wave functions can be done as
\begin{equation}
\Phi^{jm\pi} = \sum_{i=1}^N \sum_K c_i P^j_{mK} P^\pi \Psi_i,
\label{GCM}
\end{equation}
where $K$ is the projection of the angular momentum on the intrinsic axis.
The states projected on different $K$ quantum numbers are mixed.
We superpose $N$ different Slater determinants, and $N$ is 175 in the present model.
Here, $\{ \Psi_i\}$ is a set of Slater determinants with
different values \{$\pmb{R}_i$ \},
and the coefficients for the linear combination, $\{ c_i \}$, are
obtained by solving the Hill-Wheeler equation.

The projection on parity and angular momentum is performed by 
introducing the projection operators $P^j_{mK}$ and $P^\pi$,
and these are carried out numerically.
The angular momentum projection is performed using the Wigner function ${\cal D}^j_{mK}(\Omega)$
and rotation operator ${\cal R}(\Omega)$,
\begin{equation}
P^j_{mK}\Psi_i = {1 \over 8\pi^2} \int d\Omega \, {\cal D}^{j*}_{mK}(\Omega)\, {\cal R}(\Omega)\, \Psi_i,
\label{J-projection}
\end{equation}
where ${\cal R}(\Omega)$ rotates both the Gaussian center parameters $\{ \pmb R_i \}$ and the spin
part of the wave functions, and $\Omega$ stands for the Euler angles, $\alpha$, $\beta$, and $\gamma$.
We have to solve the motion of the valence neutron, which is spatially extended.
In these conditions,
we need a large number of mesh points for the Euler angles to guarantee the numerical accuracy.

The parity projection is performed by superposing another Slater determinant,
where the Gaussian center parameters are spatially inverted,
\begin{equation}
P^\pi = (1+P^r)/\sqrt{2},
\label{pi-projection}
\end{equation}
where $P^r$ is the operator which inverts the spatial coordinates 
of the Gaussian center parameters, $P^r \Psi(\{ \pmb R_i \})$ = $\Psi(\{ -\pmb R_i \})$.

\subsection{Generation of the Gaussian center parameters}
We superpose different configurations as in Eq.~(\ref{GCM}).
For this purpose, we generate many different sets of the Gaussian center parameters $\{ \pmb{R}_n\}$ ($n = 1,2,3, \ldots, 9$) in Eq.~(\ref{spwf}). We use random numbers to achieve a fast convergence of the energy based on the spirit
of the SVM~\cite{KK77,VS95} and of the antisymmetrized molecular dynamics -- superposition of selected snapshots (AMD triple-S)~\cite{PhysRevC.68.054302}.

The two-$\alpha$ cluster parts ($n = 1-8$) are introduced with a relative distance $R$,
\begin{equation}
\pmb{R}_n = -\frac{R}{2} \pmb{e}_z
\end{equation}
for $n=1-4$ and  
\begin{equation}
\pmb{R}_n = +\frac{R}{2} \pmb{e}_z
\end{equation}
for $n=5-8$, where $\pmb{e}_z$ is a unit vector along the $z$ axis.
The $R$ values are generated between 0 fm and 5 fm with a uniform distribution.

For the valence neutron ($n=9$), we have to precisely describe the wave function up to the tail region.
For the three directions ($k=x,y,z$), the Gaussian center parameter of the valence neutron
$(\pmb{R}_9)_k$ is generated using random numbers $\{r_i\}$
which are not equally distributed but have the probability  
proportional to 
$\exp(- r_i /\sigma )$,
where 
$\sigma=4$ fm is introduced.
Positive and negative values are generated with equal probability.
In this way we generate 175 
Slater determinants with different sets of Gaussian center parameters.
In the actual calculation, we prepared different sets of Gaussian center parameters 
using different random numbers and compared the results.
The energies of  the states, which are candidates for the resonances,
are almost the same, and continuum solutions
are also very similar.

In the original SVM, the selection of important basis states was performed.
On the other hand, here we employ all the basis states generated.
The selection of the basis states works well for
bound states and for narrow resonances, which are
well confined inside the interaction range. 
However, for continuum states, which are important in the present case,
the selection sometimes restricts too much the functional space.
If the number of valence nucleons increases, we eventually need a selection
of the basis states, but here, we employ all the basis states.

\subsection{$\be$ properties}

\begin{figure}[htb]
	\begin{center}
	     \epsfig{file=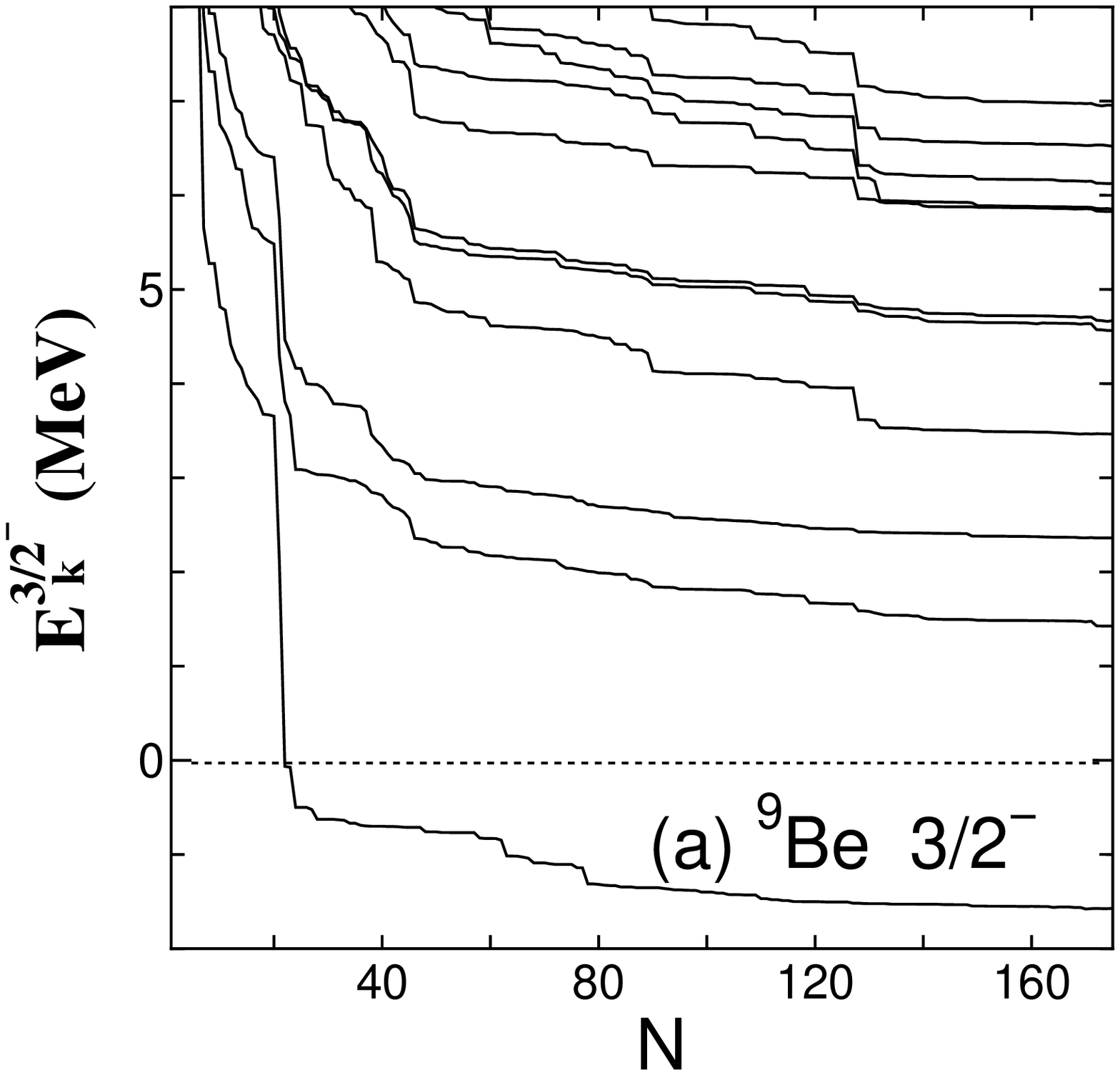,width=6.5cm}
		\epsfig{file=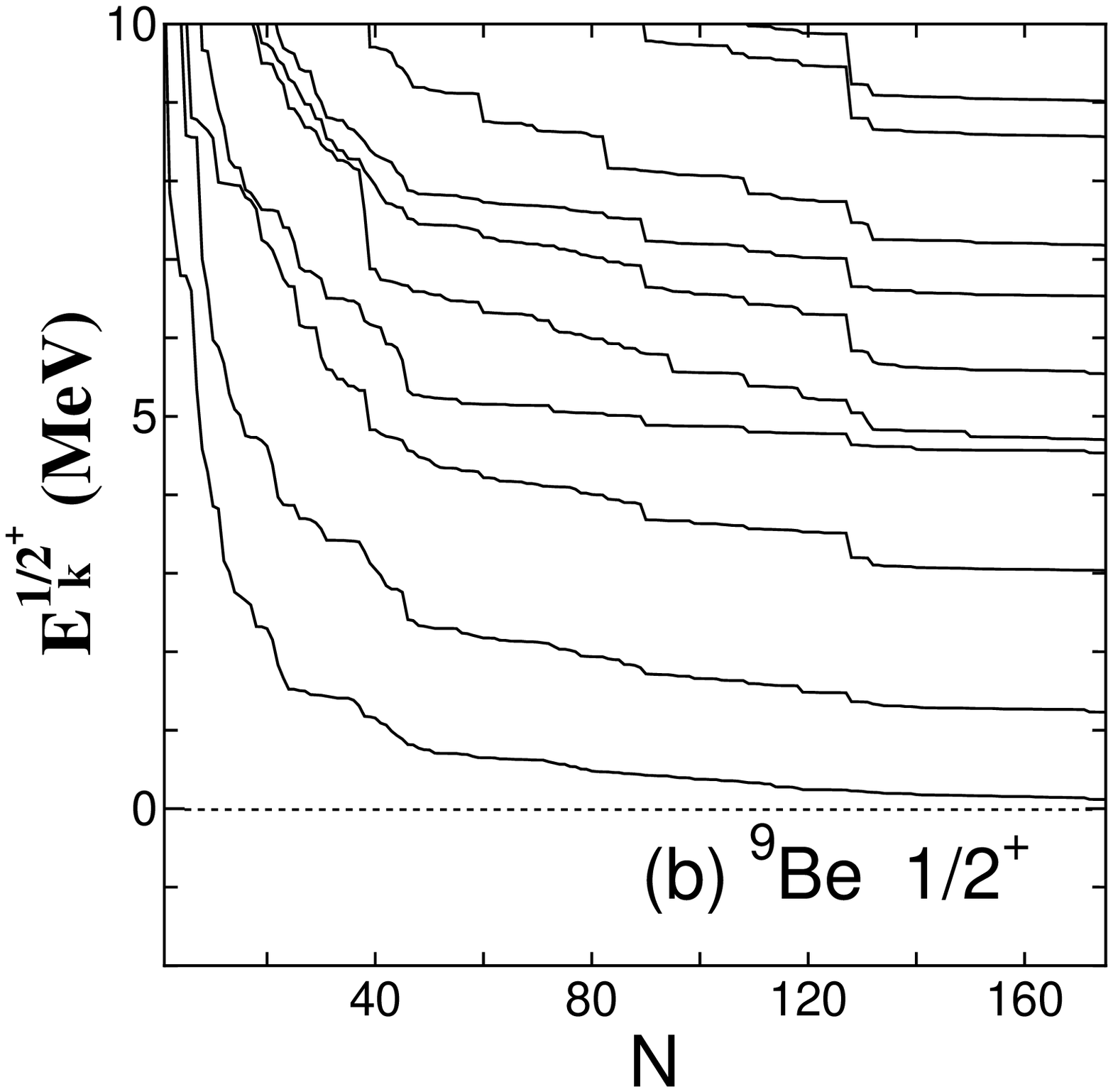,width=6.5cm}	 
		\caption{Energy convergence of the $3/2^-$ (a) and $1/2^+$ (b) states [$E^{3/2^-}_{0,k}$ and $E^{1/2^+}_{0,k}$
		in Eq.~(\ref{eq2})] 
		as a function of number of Slater determinants [$N$ in Eq.~(\ref{GCM})]. The dotted lines
		show the $\alpha$+$\alpha$+$n$ threshold.}
		\label{fig_energyconv}
	\end{center}	
\end{figure}

\begin{figure}[htb]
	\begin{center}
		\epsfig{file=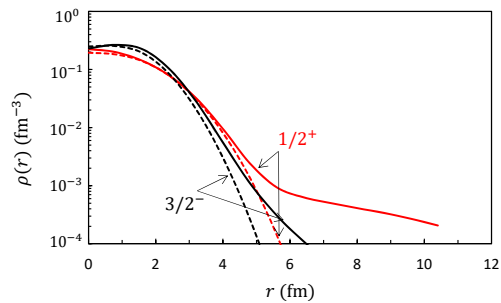,width=6.5cm}
		\caption{Neutron (solid lines) and proton (dashed lines) monopole
			densities of the $3/2^-$ and $1/2^+$ states.}
		\label{fig_dens}
	\end{center}
\end{figure}

The energy convergence of the $3/2^-$ and $1/2^+$ states 
is illustrated in Figs.~\ref{fig_energyconv} (a) and (b), respectively.
Clearly, $N=175$ provides energies close to convergence.
The monopole density distributions of the ground state and of the first excited state
are shown in Fig.~\ref{fig_dens}.
As expected, the neutron density of the $1/2^+$ state extends to large distances. This statement is true for
most pseudostates.
The root mean square matter radius of the ground state is 2.45 fm,
in excellent agreement with the experimental value \cite{THH85} ($2.45 \pm 0.01$ fm)
and previous works~\cite{PTP.57.866,PTP.61.1049,PhysRevC.54.132}.
The root mean square matter radius of the first excited state, $1/2^+_1$, is 3.83 fm.

For the electromagnetic properties,
the quadrupole moment $Q$ of the ground $3/2^-$ state
is calculated as 5.82 $e^2$fm$^2$, rather close to the experimental value \cite{TKG04} ($5.288 \pm 0.038$ $e^2$fm$^2$).
Since the charge distribution is coming only from the two $\alpha$ part, 
the $E1$ transition from the ground $3/2^-$ state to the first $1/2^+$ state
occurs as a result of  recoil effect due to the presence 
of the valence neutron.
Therefore, to properly evaluate the $B(E1)$ value, which is essential in calculating the reaction cross section,
solving the neutron wave function up to the long range region is quite important.
The present model gives $B(E1)=0.0460 \, e^2$fm$^2$, which is consistent with the observations 
($0.027(2) \sim 0.0685 \, e^2$fm$^2$~\cite{PhysRevC.82.015808}).
For E2 transitions, the $B(E2)$ value from the ground $3/2^-$ state to the $5/2^-$
state is 25.8 $e^2$fm$^4$.

\section{Microscopic CDCC formalism}
\label{sec3}
In the standard CDCC formalism, the projectile is described by a two-body \cite{Ra74} or by a 
three-body \cite{MHO04} model.  The corresponding Hamiltonian is diagonalized over a basis, 
and the eigenstates are used in an expansion of the projectile-target wave functions.  The projectile 
continuum is simulated by the positive-energy eigenvalues, referred to as pseudostates (PS).  

The total Hamiltonian of the projectile + target system is given by
\beq
H=H_0+T_R+V_{\rm int},
\label{eq1}
\eeq
where $H_0$ is the internal Hamiltonian of the projectile, $T_R$ is the kinetic energy depending 
on the relative coordinate $R$, and $V_{\rm int}$ involves optical potentials between the target 
and the constituents of the projectile.  This term depends on the internal coordinates of the 
projectile, and on the relative coordinate $R$.

The first step of the CDCC method is to diagonalize $H_0$ as mentioned in Eq.~(\ref{eq2}).
With the eigenstates $\Phi^{jm\pi}_{k}$ we define
the channel wave functions $\varphi^{JM\Pi}_{c}$ as
\beq
\varphi^{JM\Pi}_{c}=i^L
\bigl[\Phi^{j\pi}_k\otimes Y_L(\Omega_R)\bigr]^{JM},
\label{eq3b}
\eeq 
where $(J\Pi)$ are the total angular momentum and parity, and where $\Pi=\pi (-1)^L$.
The total wave function is then expanded as
\beq
\Psi^{JM\Pi}=\frac{1}{R}\sum_{c} \varphi^{JM\Pi}_{c}\, u^{J\Pi}_{c}(R),
\label{eq3}
\eeq
where index $c$ stands for $c=(j,\pi,k,L)$. 
Expansion (\ref{eq3}) assumes a spin zero for the target, but is general regarding the description of the projectile. 
 
After inserting expansion (\ref{eq3}) in the Schr\"{o}dinger equation, the radial functions $u^{J\Pi}_{c}(R)$ are 
determined from the coupled-channel system
\beq
\bigl[T_L  +E_{c}-E \bigr]  u^{J\Pi}_{c}(R) +\sum_{c'}V^{J\Pi}_{c,c'}(R) u^{J\Pi}_{c'}(R)=0,
\label{eq4}
\eeq
where the kinetic-energy operator is
\beq
T_L=-\frac{\hbar^2}{2\mu_{PT}}\left( \frac{d^2}{dR^2}-\frac{L(L+1)}{R^2}\right),
\label{eq5}
\eeq
$\mu_{PT}$ being the reduced mass of the system. In Eq.\ (\ref{eq4}), the coupling potentials are defined by
\beq
V^{J\Pi}_{c,c'}(R)= 
\langle \varphi^{JM\Pi}_{c} \vert V_{\rm int} \vert 
\varphi^{JM\Pi}_{c'} \rangle.
\label{eq6}
\eeq
The integration is performed over the internal coordinates of the projectile, and over the
relative angle $\Omega_R$. Again, Eqs.\ (\ref{eq4}-\ref{eq6}) are common to all CDCC approaches.  
The calculation of the 
coupling potentials, however, depends on the description of the projectile or, in other words, 
on the structure of the internal wave functions $\Phi^{jm\pi}_k$.

The main specificity of the microscopic CDCC is the interaction potential which reads 
\beq
V_{\rm int}(\pmb{R},\pmb{r}_i)=\sum_{i=1}^{A_p} V_{Ti}(\pmb{R}-\pmb{r}_i)
\label{eq8}
\eeq
where $\pmb{r}_i$ are the nucleon coordinates, and $V_{Ti}(\pmb{S})$ is an optical potential between 
nucleon $i$ and the target.  This potential includes the Coulomb interaction, and depends on isospin.  
The calculation of the coupling potentials (\ref{eq6}) is then performed by using a folding technique, 
which makes use of the projectile densities \cite{DH13}.

Finally, system (\ref{eq4}) is solved with the $R$-matrix method \cite{DB10,De16a} which provides the radial functions, 
and the corresponding scattering matrices.  The cross sections are deduced from 
the scattering matrices by using standard formula.

\section{Application to $\beal$ and $\bepb$ scattering }
\label{sec4}
In this section, we apply the model to two systems: $\bepb$, typical of heavy targets, 
and $\beal$, typical of light targets.  These two collisions have been studied 
experimentally \cite{WFC04,YZJ10,GAM04}, and theoretically in non-microscopic CDCC 
approaches \cite{DDC15,CRA15}.  We cover energies around the Coulomb barrier 
($E_B\approx 38.9$ MeV for $\bepb$, and $E_B\approx 8.0$ MeV for $\beal$).

\begin{figure}[htb]
	\begin{center}
		\epsfig{file=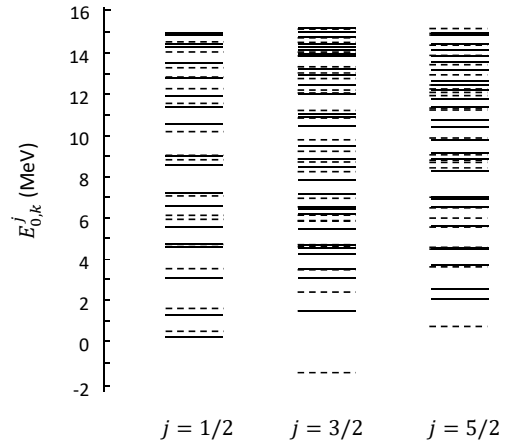,width=6.5cm}
		\caption{Pseudostate energies for $j=1/2,3/2,5/2$. The solid and dashed lines correspond
			to the positive and negative parities, respectively.}
		\label{fig_spec}
	\end{center}
\end{figure}

In Fig.~\ref{fig_spec}, we show the $\be$ pseudostate energies for angular momenta 
$j=1/2,3/2,5/2$. Positive-parity states are indicated by solid lines, and
negative-parity states by dashed lines. In addition to the $3/2^-$ ground state, the model also
reproduces the low-energy $1/2^+$ and $5/2^-$ resonances. All other states are approximations of the 
$\alpha+\alpha+n$ continuum, and do not correspond to physical states.

For the neutron-target optical potential, we take the local potential of Koning and Delaroche \cite{KD03} at a neutron energy $E_n=E_{\rm lab}/9$.  The proton-target interaction only contains the Coulomb term.  
In all cases, the proton energy $E_p=E_{\rm lab}/9$ is much lower than the Coulomb barrier of the 
$p$+target system, and the corresponding cross sections are purely Rutherford.  We take a truncation 
energy $\emax=15$ MeV, and a maximum angular momentum of $\jmax=7/2$.  Several tests have been done 
to check the stability of the cross sections against these parameters.

\begin{figure}[htb]
	\begin{center}
		\epsfig{file=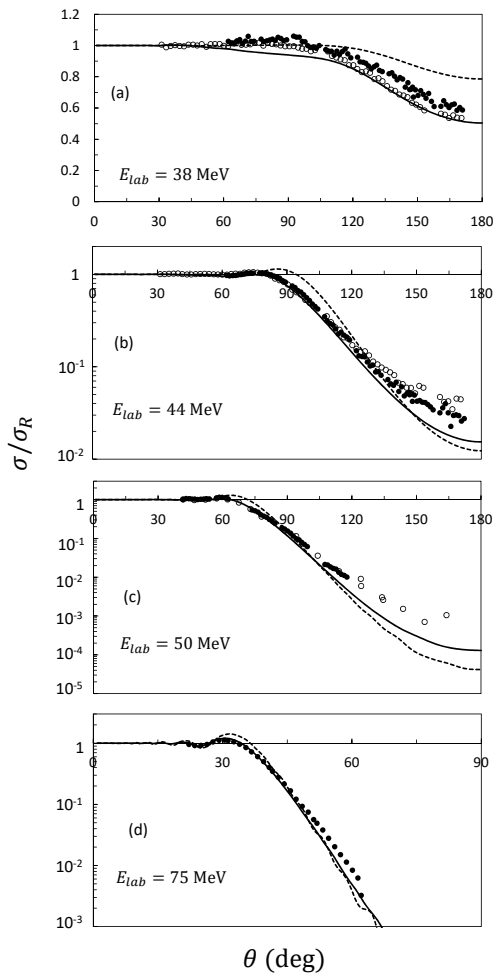,width=6.5cm}
		\caption{$\bepb$ elastic cross sections (divided by the Rutherford cross section)
			at different $\be$ laboratory energies. Dashed lines:
			single-channel calculations limited to the $\be$ ground state; 
			solid lines: full calculations. 
			The experimental data are taken from
			Ref.~\cite{WFC04} (filled circles) and Ref.~\cite{YZJ10} (open circles).}
		\label{fig_pb208}
	\end{center}
\end{figure}

The $\bepb$ cross sections are presented in Fig.\ \ref{fig_pb208}, with the data of Refs.\ \cite{WFC04,GAM04}.  
We have selected four typical energies, $E_{\rm lab}=38, 44, 50$ and 75 MeV.  We compare the full CDCC 
calculation with the single-channel approximation, i.e.\ by neglecting $\be$ breakup. At all energies, 
we have a fair agreement with the data when breakup is included.  As found in Refs.\ \cite{DDC15,CRA15}, the single-channel approximation significantly deviates from the data. 

\begin{figure}[htb]
	\begin{center}
		\epsfig{file=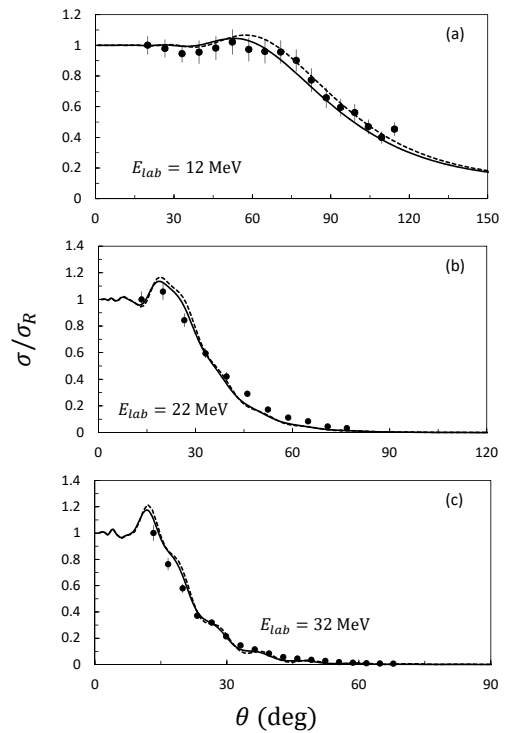,width=6.5cm}
		\caption{$\beal$ elastic cross sections (see caption to Fig.~\ref{fig_pb208}). The experimental data are taken from
			Ref.~\cite{WFC04}}
		\label{fig_al27}
	\end{center}
\end{figure}

An example with a light target, $\beal$, is shown in Fig.\ \ref{fig_al27}.  Again, the agreement with 
the experimental data is quite good, considering that there is no free parameter in the model.  
For light systems, however, the role of the breakup channels is minor.  This was already found in a 
non-microscopic CDCC analysis \cite{CRA15}.

\section{Conclusion}
\label{sec5}
In this work, we have applied the SVM to $\be$ wave functions, with the aim of performing CDCC scattering
calculations. Elastic scattering is one of the main tools to investigate exotic nuclei, and developing accurate reaction models is a challenge for theory. Owing to their low breakup threshold, exotic nuclei can be easily broken up,
and this property must be taken into account in scattering calculations.

The present description of $\be$ is based on a microscopic multicluster $\alpha+\alpha+n$ model. The wave functions depend on all nucleon coordinates, and are fully antisymmetric. The cluster approximation is used
to solve the Schr\"odinger equation associated with $\be$. We use the SVM to optimize the basis functions, where the
distances between the $\alpha$ particles, and between their c.m. and the additional neutron are parameters. Optimizing the parameter set is crucial when the number of parameters increases.

We have applied the model to $\bepb$ and $\beal$ elastic scattering at various energies around the Coulomb
barrier. The only input is the nucleon-target optical potential, which is well known over a wide range
of target masses and of nucleon energies. In both cases, we find a fair agreement with the experimental data.

Our goal for the future is to investigate reactions involving heavier Be isotopes, where the number of
degrees of freedom in the basis functions is larger. The present application to $\be$ shows that the method is promising, and that reactions involving $^{10}$Be or $^{11}$Be should be feasible in a near future.
	
\section*{Acknowledgments}
P. D. is Directeur de Recherches of F.R.S.-FNRS, Belgium. 
N. I. thanks the computer facility of 
Yukawa Institute for Theoretical Physics,
Kyoto University and JSPS KAKENHI Grant Number 17K05440.


%

\end{document}